\documentclass[doublespacing]{elsart}

\usepackage{graphicx}
\usepackage{color}
\def\Sit{Si$_3$N$_4$}
\def\TiOd{TiO$_2$}
\def\SnOd{SnO$_2$}
\def\Jm2{Jm$^{-2}$}

\begin{document}

\begin{frontmatter}



\title{Asymmetric Silver to Oxide Adhesion in Multilayers Deposited on Glass by Sputtering}

\author[label1]{E. Barthel}
\ead{etienne.barthel@saint-gobain.com},
\author[label1]{O. Kerjan},
\author[label1]{P. Nael}and
\author[label2]{N. Nadaud}
\address[label1]{Unit\'e Mixte CNRS/Saint-Gobain " Surface du Verre et Interface ",
Saint-Gobain Recherche, BP 135, F-93303, Aubervilliers, Cedex,
France.}
\address[label2]{Saint-Gobain Recherche, BP 135, F-93303, Aubervilliers, Cedex, France.}

\author{}

\address{}

\begin{abstract}
We have developed a wedge-loaded double-cantilever beam adhesion
measurement set-up for thin films deposited on glass by
sputtering. The test is described in details. Results on the
Glass/sublayer/Ag/ZnO multilayer provide evidence that \SnOd\ or
\TiOd\ perform better than ZnO as a sublayer. Then however,
rupture within the multilayer shifts to the upper Ag/ZnO
interface. The latter is shown to be tougher than the lower ZnO/Ag
interface, an asymmetry due to non-equilibrium interfacial
structures.
\end{abstract}

\end{frontmatter}

\section{Introduction}
Thin film multilayers deposited on glass are widely used for flat
optical, photoelectric and electrochromic devices. Typical
applications are Infra-Red filters against heat-generating solar
radiations (solar control), solar cells for electrical power
generation or voltage-controlled devices for tunable absorption in
the visible light frequency range. For instance, in the present
paper, we consider a ZnO/Ag/ZnO sandwich, with respective layer
thicknesses of 20, 10 and 20 nm. This stack stands as a prototype
both for solar control coatings~\cite{Ando} and for solar cell
electrodes~\cite{Ray}.

\textcolor{red}{In the metal/oxide adhesion literature, the metals
are classified according to their ability to react with the oxide
and form an interphase. Silver, along with gold and platinum belongs
to the group of non-reactive metals~\cite{Stoneham} which form an
abrupt interface with all oxides, without interphase. The resulting
adhesion energies are low.}
In many applications, however, adhesion is a crucial issue. This
is the case for instance when mechanical strength is required for
further processing or for integration in complex systems. Scratch
resistance is also a general concern either during process or
during service.

Assessment of the adhesion of such thin films is therefore
necessary. Although more academic approaches such as high
temperature sessile drop methods are useful to grasp some of the
underlying mechanisms, they \textcolor{red}{cannot} take into
account all the specifics of adhesion within a multilayer: such
characteristics as stoechiometry or structure are largely
dependent upon the deposition conditions. Therefore, for thin
films, one of the primary requirements is to measure the adhesion
directly on the system under study.

\begin{figure}
\begin{center}
\includegraphics[width=3.25in]{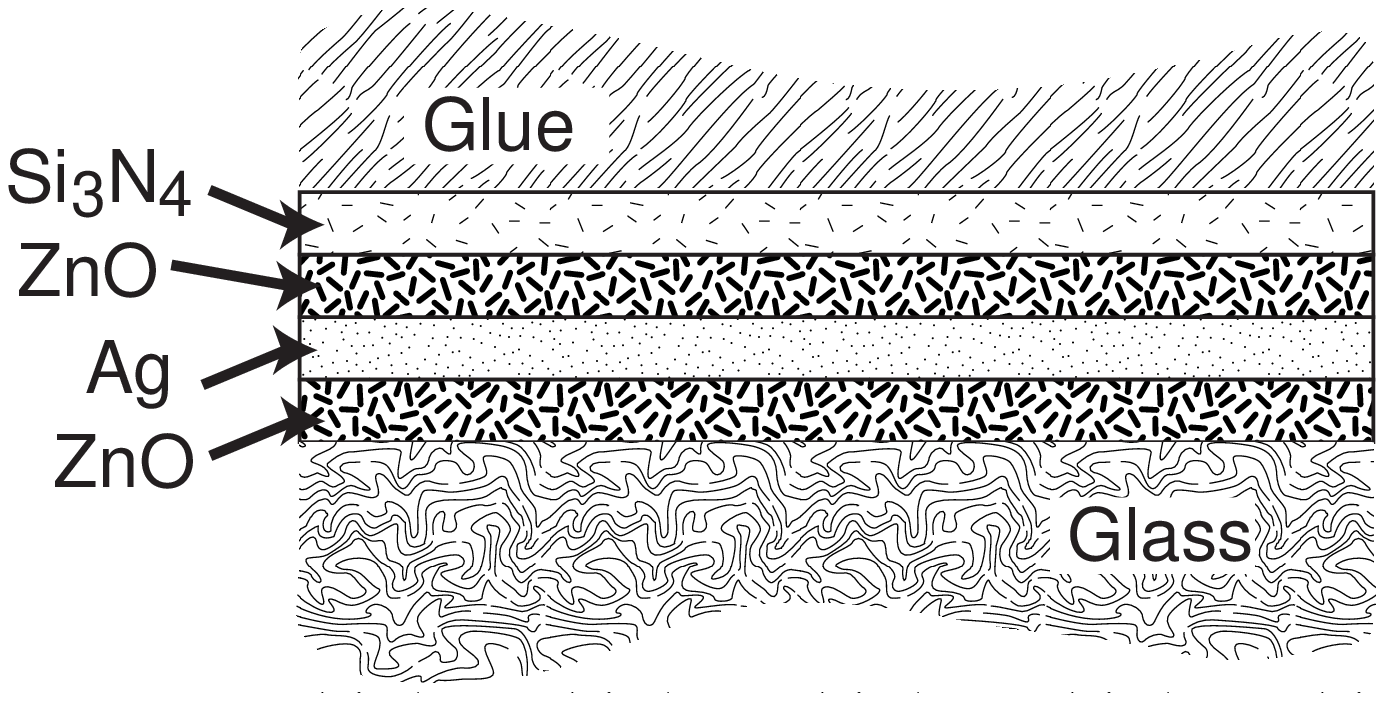}
\caption{}\label{layers}
\end{center}
\end{figure}

For that purpose, we have developed a set-up to measure the
adhesion energy of multilayer films in the tens of nanometer
thickness range deposited by RF sputtering on thick glass
substrates. In the present paper, the effect of modifications or
substitutions of the lower ZnO layer in ZnO/Ag/ZnO stacks
(Fig.~\ref{layers}) was investigated. The relative performance of
\Sit, ZnO and \SnOd\ are compared. The crack path selection
mechanism is discussed. Through the adhesion measured in this way,
we also evidence more complex phenomena:
\textcolor{red}{in particular, we show directly that deposition of
the metal on top of the oxide does not result in the same work of
adhesion as deposition of the same oxide on top of the same metal.
The work of adhesion is also shown to depend upon the next nearest
layers, {\it i.e.} layers which are not directly adjacent to the
interface of rupture.}

\section{Experimental details}

\subsection{The layers}

The systems studied here were deposited on the glass substrate by
magnetron sputtering using an Alcatel Lina350 in-line sputtering
system. The typical stack (Fig.~\ref{layers}) is Glass/sublayer
(20)/Ag (10)/ZnO (20) where the figure in brackets is the layer
thickness in nanometers. The stack is terminated by a \Sit\ layer of
thickness varying between 4 and 35 nm. A number of sublayers such as
ZnO, \SnOd\ and \Sit\ were tested. ZnO and \SnOd\ were obtained by
reactive sputtering of respectively Zn and Sn planar targets  using
argon and oxygen as primary sputtering and reactive gases.
\textcolor{red}{All the layers within the multilayer are deposited
in-line, without breaking the vacuum, with a pre-sputter time of 3
minutes.}
\Sit\ was obtained by reactive sputtering of a polycrystalline Si
target using argon and nitrogen as primary sputtering and reactive
gases. Ag was obtained by sputtering an Ag planar target using
argon as sputtering gas. For all the experimental runs, the
background pressure before
\textcolor{red}{deposition}
was about 7.10$^{-7}$ mbar and the total sputtering pressure was
8.10$^{-3}$ mbar.
The cathodic power applied to the targets \textcolor{red}{was 200
and 2000 W} for Ag and dielectric materials, respectively.
For reactive sputtering, oxygen and nitrogen partial pressures
were adapted for a full oxidation or nitridation of the \SnOd, ZnO
or \Sit, respectively.
All substrates were cleaned by hot demineralized water and
\textcolor{red}{mechanical brushing}.

\subsection{Adhesion energy measurements - The Double Cantilever Beam test}
The literature on thin film adhesion energy measurements is
wide~\cite{Hou,Drory,Evans,Mittal,Volinsky}. This results from the
fact that many measurement set-ups are specific to a given
film-substrate combination. The applicability of a given method
will depend, among other parameters, upon the thickness of the
layer, the respective mechanical properties of the layer and the
substrate, the relevant crack velocity.

For thin films, a number of experimental set-ups, which are easy
to implement, such as the scratch test or the pull-out test,
return qualitative rather than quantitative results. For more
quantitative measurements, it is necessary to apply the mechanical
stress with the help of a backing of some sort. Here, the backing
is made out of glass: in our cleavage set-up, two glass plates,
one of them bearing the multilayer, are glued together
(Fig.~\ref{sampleopen} a). Cleavage of such a sandwich is a
complex physico-mechanical process
\textcolor{red}{which involves dissipation through irreversible
processes.}
\textcolor{red}{In such cases, the interfacial toughness $G$ ({\it
i.e.} the measured work of adhesion) can often be expressed as
\begin{equation}
G=w\phi
\end{equation}
where $w$ is the thermodynamic work of adhesion for reversible
surface separation and $\phi$ expresses the enhancement of the work
of adhesion due to irreversible processes. Extraction of the
thermodynamic work of adhesion from the interfacial toughness is
seldom straightforward~\cite{Schapery,Wei}. This extraction will not
be attempted here. However, we will assume that the enhancement
factor $\phi$ is constant and the test will only be used to compare
thermodynamic work of adhesion through the interfacial toughness.}

\subsubsection{Description of the test}
A glass backing identical to the 2.3 mm float glass substrate is
glued onto the thin film.
\textcolor{red}{Typical samples are 70 mm long and 50 mm wide.}
The glue is Epotecny 505, a two-components epoxy which was
prepared as specified by the manufacturer and cured at
80$^{\circ}$C for 45 min. The glue layer is about
\textcolor{red}{25$\pm 5$~$\mu$m thick}, with a 2 GPa Young's
modulus.
\textcolor{red}{The multilayers are stable to at least 300
$^\circ$~C, in particular without silver dewetting, suggesting they
are unaffected by the thermal treatment. No trace of glue diffusion
within the multilayer was observed by X-ray photoelectron
spectroscopy (XPS).}

\begin{figure}
\begin{center}
\includegraphics[width=3.25in]{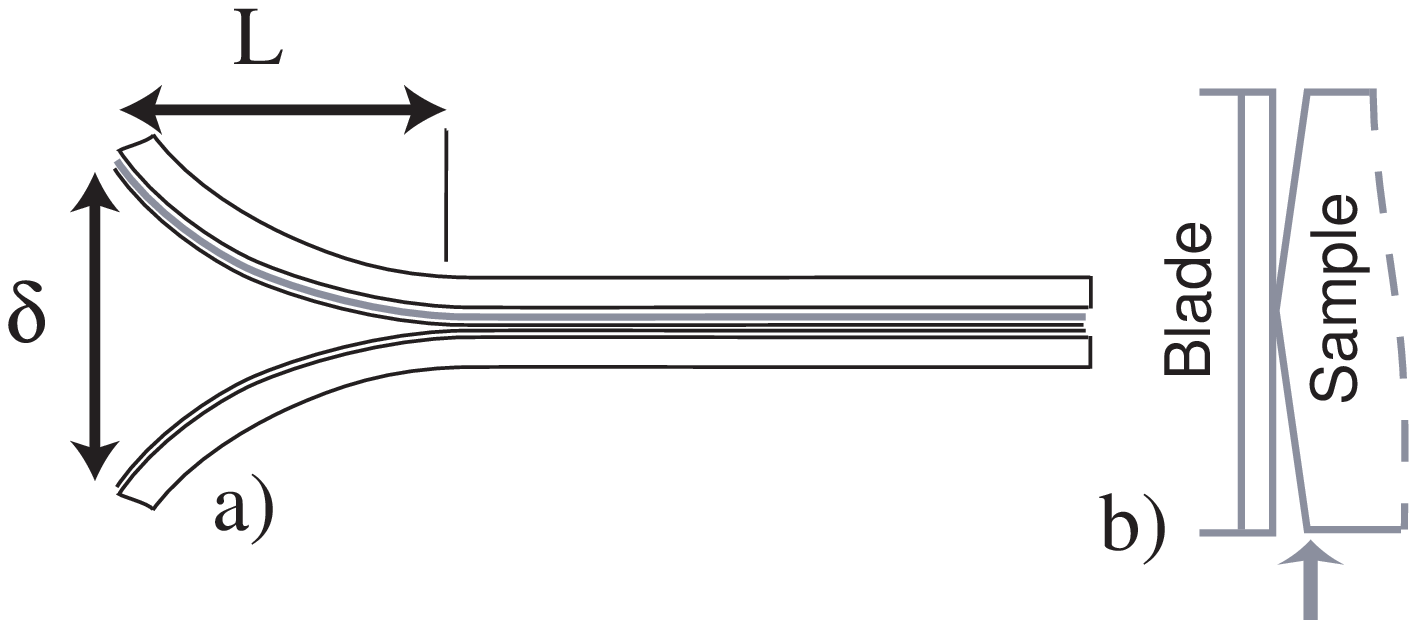}
\caption{}\label{sampleopen}
\end{center}
\end{figure}

The cleavage of the sample is obtained by the gradual introduction
of a bevelled blade. The blade is mounted on an electric jack, which
allows for precise positioning. In this way, the opening
displacement $\delta$ of the arms of the DCB sample can be
controlled (Figure~\ref{sampleopen} a): the typical opening lies in
the 30 $\mu$m to 250 $\mu$m range. It is measured by a high
magnification camera (Figure~\ref{sampleopen} b).

In this way, the test is conducted at fixed grip and is
mechanically stable, {\it i.e.} catastrophic rupture is avoided
and the crack length can be increased in a controlled manner. As a
result, the following experimental procedure is used: for a given
sample, the opening is gradually increased and for each value of
the opening, the crack length $L$ -- typically 2 to 4 cm -- is
measured with a transparent ruler. Due to viscous relaxation in
the glue, it takes a few minutes for the crack position to
stabilize. A typical waiting time of 15 min is allowed for between
each opening increment. The test is conducted at ambient air.

\subsubsection{Crack propagation control}
Careful sample preparation is crucial to prevent surface flaws
from propagating into the glass and ruining the sample. For that
purpose, after curing the glue, the sample is carefully re-cut
into its final rectangular shape so as to remove glue spill-outs.
At this stage, flaws on the glass edges should be avoided or
suppressed by polishing so as not to compromise the overall sample
strength.
 Of course, crack initiation must take place within the
multilayer. For that purpose, release layers may be avoided if in
the final re-cutting step one of the small sides of the sample is
cut into a pointed end (Fig.~\ref{sampleopen} b); pre-cracking is
then achieved by pressing the blade onto the glue joint at the tip
edge. In the systems studied here, this results in crack
initiation: wherever it started, this initial crack soon
propagates into the multilayer where it stabilizes at a definite
interface. With this crack initiation procedure, a broader range
of samples is suitable for adhesion measurements, including those
where a release layer cannot be provided for.

\subsubsection{Locus of failure}
\begin{figure}
\begin{center}
\includegraphics[width=3.25in]{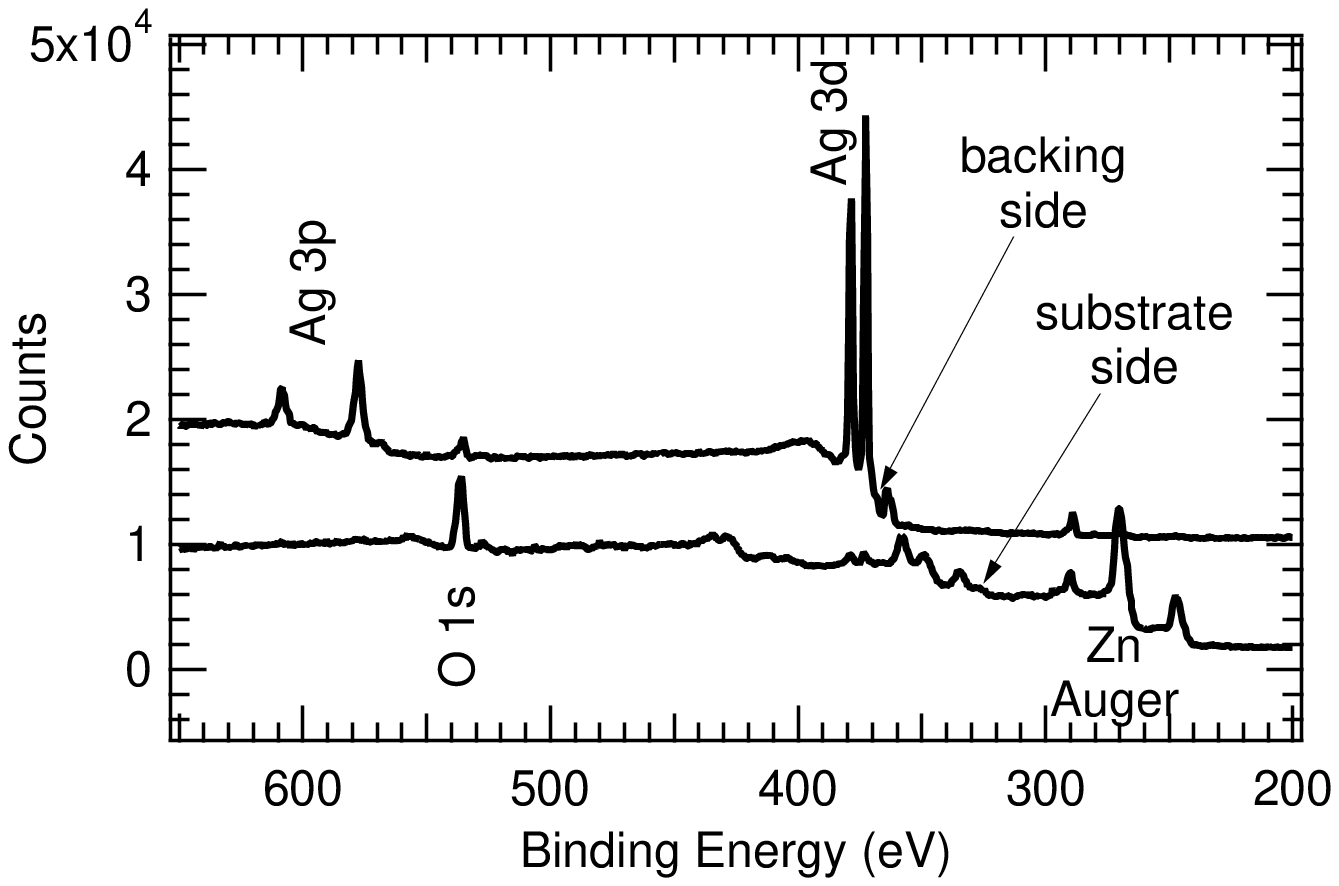}
\caption{}\label{XPS}
\end{center}
\end{figure}

Identification of the locus of failure is of primary importance for
attribution of the adhesion energy measured to a specific interface
and also for assessment of the nature of the crack propagation. It
is achieved by XPS \textcolor{red}{(CLam 2, Fisons Instruments, Mg
K$\alpha$)} of both cleavage surfaces. In the present systems,
perfectly interfacial ruptures at a definite interface within the
multilayer have always been found, with negligible material transfer
on the opposite surface, \textcolor{red}{as illustrated on
Fig.~\ref{XPS}}. This is confirmed by the very low roughnesses
\textcolor{red}{(less than 1 nm RMS over 1 $\mu$m$^2$)} measured
after cleavage by Atomic Force Microscopy (AFM).

\subsubsection{Data \textcolor{red}{interpretation}}
There are two steps in the data \textcolor{red}{interpretation}.
The first one is to calculate the \textcolor{red}{interfacial
toughness} from the cleavage data. This depends on the mechanics
of the glass arms, is relatively straightforward and is detailed
in this section. The next step is to make a connection between the
mechanics of the test -- {\it i.e.} geometry, material properties
and \textcolor{red}{interfacial toughness} -- and the decohesion
process at the interface. This step involves a mechanical
description at the scale of the glue layer and is much more
involved. It will not be attempted here in any detail: we will
only argue later in the discussion that a simple monotonic
relation between \textcolor{red}{interfacial toughness} and
\textcolor{red}{thermodynamic} work of adhesion holds and that
rupture occurs at the weakest interface.

The \textcolor{red}{interfacial toughness} is derived from the
data through the standard augmented beam model by
Kanninen~\cite{Kanninen}. However the elastic foundation
contribution is actually negligible even when the glue layer is
taken into account~\cite{Penado}, so that the simplest beam
theory~\cite{Obreimov} would be adequate. According to the
augmented beam model
\begin{equation}\label{G}
  G=\frac{3Eh^3\delta^2}{16(L+0.6h)^4}
\end{equation}
where $\delta$ and $L$ are the crack opening and crack length, and
$E$ and $h$ the beam Young's modulus and thickness, and $G$ is the
energy release rate. Indeed, beam bending stores elastic energy.
The energy release rate $G$ is the amount of elastic energy
released upon incremental crack advance. At mechanical
equilibrium, it is equal to the toughness of the adhesive joint.
In turn, there is a relation between this interfacial toughness
and and the \textcolor{red}{thermodynamic} work of adhesion of the
interface of rupture. This relation will be discussed below.

\section{Results}
\begin{figure}
\begin{center}
\includegraphics[width=3.25in]{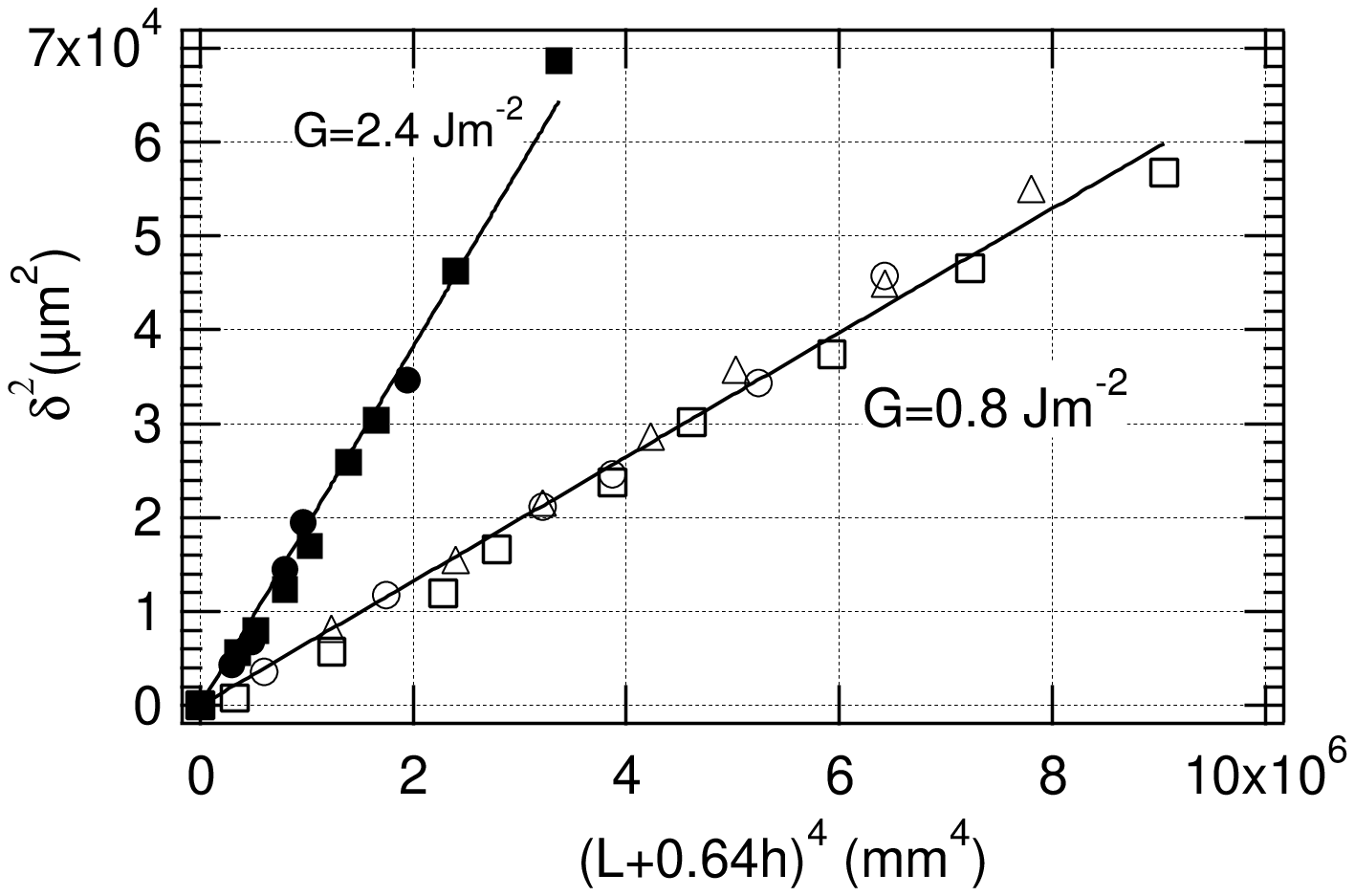}
\caption{}\label{Plot}
\end{center}
\end{figure}

Typical results plotted according to Eq.~\ref{G} are displayed in
Fig.~\ref{Plot}. The linear plots confirm the applicability of the
Kanninen model. Good \textcolor{red}{repeatability} from sample to
sample is also evidenced by the \textcolor{red}{superimposition}
 of the data. The results for the various sublayer
substitutions are summarized in the Table. The interface of
rupture as identified by XPS is denoted by a // in the stack
description.

A \Sit\ sublayer (system 1) results in the weakest joint. Rupture
occurs between \Sit\ and \textcolor{red}{silver}. ZnO (system 3)
improves on \Sit, but ZnO on \Sit\ (system 2) is not as good as
bare ZnO. In these last two cases, rupture is located between the
{\em lower} ZnO layer and silver. \TiOd\ (system 4) and above all
\SnOd\ (system 5) perform significantly better than ZnO. Then,
however cohesion fails between \textcolor{red}{silver} and the
{\em upper} ZnO layer.

\section{Discussion}
\subsection{Relation between \textcolor{red}{interfacial toughness} and \textcolor{red}{thermodynamic} work of adhesion}
Comparison of the interfacial toughness values measured here with
typical metal/oxide adhesion energies measured by the sessile drop
technique suggest that our values overestimate the adhesion
energy. For instance, a representative value for the adhesion
energy of \textcolor{red}{silver} on a large gap oxide like
sapphire is 0.34 \Jm2~\cite{Sotiropoulou}. Since
\textcolor{red}{silver} obviously dewets on \Sit, we would expect
the adhesion energy in this case to be lower, of the order of 0.15
- 0.3 \Jm2.

As expected, the difference between the \textcolor{red}{interfacial
toughness} and the thermodynamic work of adhesion values demonstrate
that the elastic energy release rate measured includes additional
effects. This is where the details of the mechanics of the system at
the local scale have to be considered. \textcolor{red}{A first kind
of effects result from residual stresses in the layers. They are
expected to be of small magnitude for such thin layers in a cleavage
geometry (appendix \ref{relaxation})}. To explain an enhancement of
the adhesion, additional mechanical dissipation processes are
invoked. Since the rest of the system is essentially brittle (glass)
or negligibly thin (\textcolor{red}{silver}), this dissipation most
likely takes place within the glue layer. A sizeable viscoelastic
contribution is ruled out because the tests are conducted at
virtually zero crack tip velocity. The source of dissipation is
plastic deformation in the glue: more details may be found in
appendix~\ref{dissip}. In brief, we assume the plastic dissipation
takes the form of an enhancement factor which is more or less
constant in the adhesion energy range considered here.

\subsection{Interfaces within the multilayers - Adhesion}

\subsubsection{The Lower Interface}

For non reactive metals like \textcolor{red}{silver}, the adhesion
to oxides is still not well understood. The difficulty arises from
the fact that several effects may contribute to the -- low --
final value. A short and recent review may be found in
ref.~\cite{Campbell}. General trends have been experimentally
identified such as decreasing adhesion with increasing oxide
gap~\cite{Li} or increasing adhesion with increasing enthalpies of
mixing~\cite{Sangiorgi}.

Along these lines, the reduction of adhesion when \Sit\ is substituted for ZnO can be rationalized at least in two ways: a larger gap and a smaller metal/nitrogen than metal/oxygen affinity.

The reduction of adhesion when the ZnO layer is deposited on top
of a \Sit\ sublayer signals intrinsic multilayer issues: the
nature or structure of the next nearest layer influences the
interface. Indeed, the \Sit\ layers are significantly rougher than
the bare glass substrate. Along with surface chemistry, this forms
a possible reason to alter the growth of the subsequent ZnO layer,
which then offers a lower interfacial energy to
\textcolor{red}{silver}.

\subsection{Crack path selection}\label{CrackPath}

Our data show that \SnOd, and also \TiOd, provide a more adhesive
substrate to \textcolor{red}{silver} than ZnO. The latter is
hygroscopic so that environmental conditions and water assisted
corrosion may be crucial here. Moreover, the crack path changes
when ZnO is replaced by one of theses oxides. This means that the
toughness then measured is actually the toughness of the upper
Ag/ZnO interface.

Before comparing the relative interfacial toughnesses, the crack
path selection mechanism should be discussed. Crack path selection
is a non trivial issue in the case of the bi-material sandwich we
use because crack propagation on one side of the glue layer breaks
the symmetry so that the details of the stress field at the crack
tip (mode mixity) may induce crack deviations~\cite{Wang}.

Elastic calculations~\cite{Fleck,Ritchie}, which are valid in the
present case provided the plastic zone is not too large, suggest
that the glass/epoxy system lies on the borderline between stable
growth within the layer and propagation close to the interface. In
the present case, we may also rely on the aluminum/epoxy data:
indeed the elastic properties of glass and Aluminum are very
close. \textcolor{red}{In fact, in the Aluminum/epoxy sandwiches,
calculations predict a small negative value of the phase angle
(around -13$\,^{\circ}$)~\cite{Wang} while} experimentally it is
observed that with a symetrical loading, rupture occurs in the
middle of the epoxy layer, despite the higher
toughness~\cite{Wang}. To sum up, in Fig.~\ref{layers}, we expect
a weak tendency for crack deviation upwards, towards the more
compliant part.

However, rupture at the interface is observed here. This is likely
to occur if the interfacial energy is substantially smaller than
the cohesion energy of the glue (a few hundred J/m$^2$). In our
case, it is this energy criterion \textcolor{red}{which} primarily
controls the crack path.

However, the crack deviation mechanism should be kept in mind when
the question of the crack path selection {\em within} the
multilayer arises. An interesting perspective in the present
experimental set-up is set by the possibility of tuning the
asymmetry of the stress field at the crack tip by the asymmetry of
the glass arms of the DCB. Hopefully, the crack propagation, and
thus the interface of rupture within the multilayer could be
controlled in this way.

\subsubsection{The Upper Interface: Asymmetry}
In brief, we have observed that: 1) on the ZnO/Ag/ZnO stack, the
crack always propagates at the lower interface, although the mode
mixity tends to drive it upwards, to the glue layer; 2) when the
crack propagates at the upper Ag/ZnO interface, the
\textcolor{red}{interfacial toughness} is larger than when it
propagates at the lower.

The simplest scenario is that the {\it
\textcolor{red}{thermodynamic work of adhesion}} of the upper
interface is actually larger than the lower, thus driving the
crack to the lower interface. An alternative explanation is that
the increased adhesion energy at the upper interface is due to an
increase in plastic dissipation because of the reduced thickness
of the top elastic layer. However, this is not the case since the
same behaviour is observed with thicker upper layers, increasing
the top \Sit\ layer to 35 nm.

In both cases, we conclude that the \textcolor{red}{thermodynamic
work of adhesion} for the \SnOd/Ag and \TiOd/Ag interfaces is at
least as large as for the Ag/ZnO interface. This means that
substituting \SnOd\ or \TiOd\ for ZnO results in at least about a
50\% increase. \textcolor{red}{This is not due to roughness
effects because the \SnOd\ and ZnO layers have similar roughness.
However, it is consistent with the observation that non-reactive
liquid metals wet \TiOd\ but not ZnO~\cite{Stoneham}}.

The asymmetry between the ZnO/Ag and Ag/ZnO adhesion is surprising
at first sight because the {\it equilibrium} expression for the
adhesion energy, the Young's equation, is symmetric with respect
to the materials involved. However, during the growth of the
layer, equilibrium is not reached. This is illustrated by the high
temperature dewetting of \textcolor{red}{silver} films on oxide
substrates. Several factors like adatom diffusion or spreading
pressure will differ for Ag on ZnO or ZnO on Ag deposition. The
resulting structure of the interface and therefore the adhesion
energies will then be different. A similar case of asymmetric
interfaces appear in the Mo/Si multilayers developped for EUV
reflective optics~\cite{Yulin}.
\section{Conclusion}
We have applied the wedge controlled DCB test to thin multilayers
deposited on glass by sputtering. A specific sample preparation
and experimental procedure allow for reliable measurements. In
terms of adhesion of the \textcolor{red}{silver} layer, the effect
of the nature of the sublayer has been quantified:\Sit, ZnO,
\TiOd\ and \SnOd\ exhibit increasing performances. For the last
two, the crack path switches to the upper Ag/ZnO interface, the
adhesion energy of which is actually the measured quantity. We
provide evidence that this upper Ag/ZnO is stronger than the lower
ZnO/Ag. This asymmetry is due to the non equilibrium
configurations of these otherwise identical interfaces.
Next-nearest layers also play a role in the adhesion.
\section{Acknowledgements}
We \textcolor{red}{thank} J. Jupille for interesting discussions.

\appendix

\section{The mechanics of the test}
\subsection{Intrinsic stresses relaxation}\label{relaxation}
Such layers as ZnO or \Sit\ may store significant residual
stresses. \textcolor{red}{Their contribution to the thermodynamic
work of adhesion will depend upon stress sign and test geometry}.
However, it is expected that in such confined geometries, minute
film deformations are allowed, so that very little of these
stresses relax upon crack propagation~\cite{Dauskardt}. As a
result, residual stress contributions to the measured energy
release rates will not be considered here.

\subsection{Plastic dissipation}\label{dissip}
\begin{figure}
\begin{center}
\includegraphics[width=3.25in]{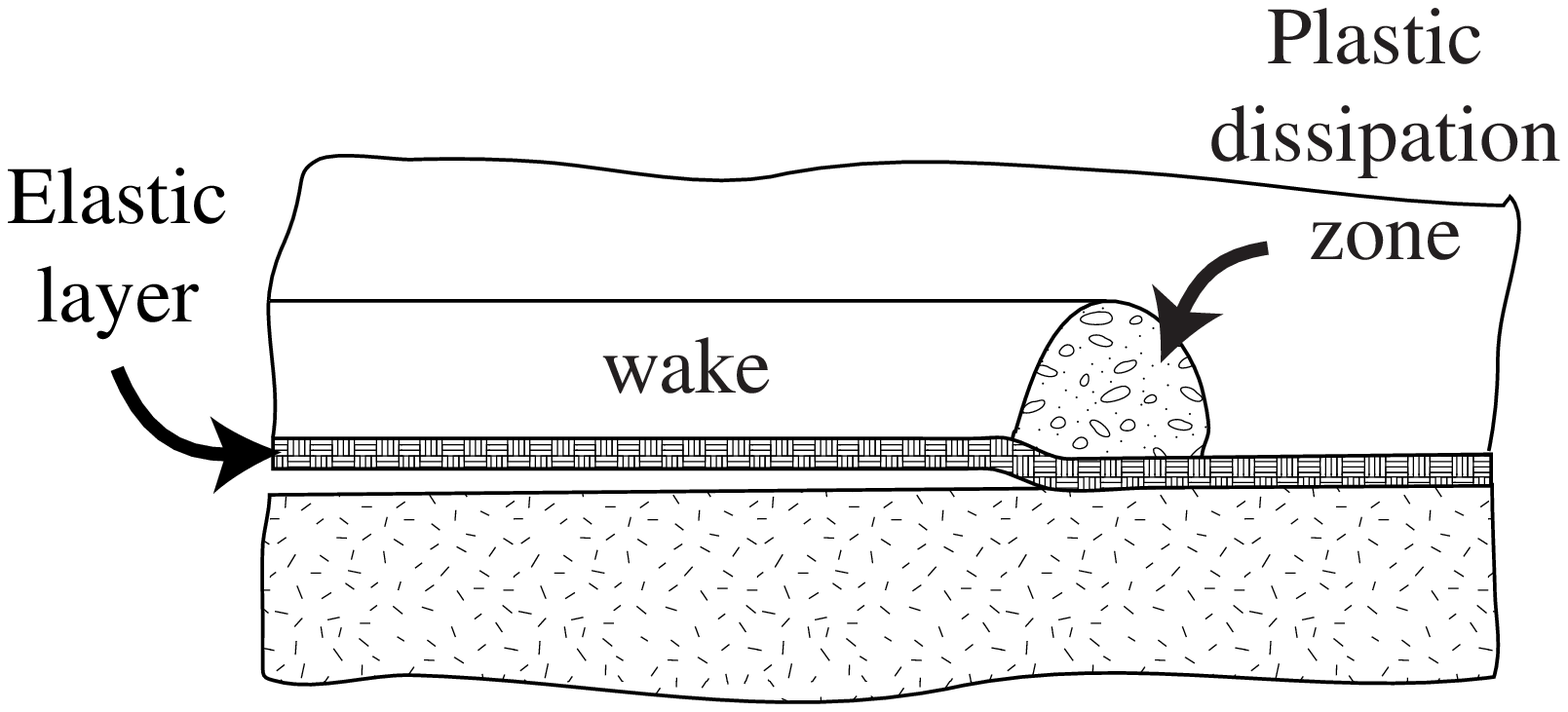}
\caption{}\label{Plast}
\end{center}
\end{figure}

Assuming that the glue and glass Young's modulus are respectively
2 and 70 GPa, and that the glue yield stress is in the 30 to 100
MPa range, a plastic zone size of roughly 1 - 10 $\mu$m can be
estimated~\cite{Wei}. This result means that confinement of the
plastic zone by the glass backing is not essential. This is
essentially due to the low adhesion energies encountered in the
present systems. As a result, the thickness of the glue layer is
not a first order parameter. It would be tempting to try and
calculate the enhancement factor relating
\textcolor{red}{interfacial toughness} and
\textcolor{red}{thermodynamic work of adhesion}. Unfortunately,
this would require extensive finite element analysis, taking into
account the complex tensile behaviour of the epoxy glue and the
thickness of the residual layer between fracture plane and glue
(Fig.~\ref{Plast}). It is therefore possible, but was not
undertaken here.

We only assume that the larger the \textcolor{red}{thermodynamic
work of adhesion}, the larger the \textcolor{red}{interfacial
toughness}. Thus, by comparing toughnesses, we compare
\textcolor{red}{thermodynamic works of adhesion} indirectly.





\newpage
\section*{Captions}
\subsection*{Table}
Table: \textcolor{red}{interfacial toughness} for different
multilayers as measured with the wedge-loaded DCB test with a
glued glass backing. The interface of rupture is denoted by the //
sign.
\subsection*{Figures}
Fig. 1: Schematics of a typical multilayer. Layer thicknesses are in
the 10-20 nm range. In this example, the sublayer (between the glass
substrate and the silver layer) is ZnO.

Fig. 2: Schematics of the cleavage set-up: a) side view: the wedge
controled opening $\delta$ and the crack length $L$ are measured
and the \textcolor{red}{interfacial toughness} calculated with the
Kanninen model (Eq.~\ref{G}); b) top view: end cut for crack
initiation, \textcolor{red}{showing also the position of the
blade; the arrow indicates the position where the opening $\delta$
is masured.}

\textcolor{red}{Fig. 3: Typical XPS spectra of the cleaved surfaces
for a substrate/ZnO/Ag/ZnO/backing stack. The rupture, which
occurred between the silver layer and the ZnO on the substrate side
(glass), is almost perfectly interfacial, with negligible material
transfert.}

Fig. 4: Typical data plot according to the Kanninen model.
\textcolor{red}{The slope is proportional to the interfacial
toughness $G$ (Eq;~\ref{G}).} The points where collected from
respectively two and three samples for the high adhesion (filled
symbols) and low adhesion (empty) samples.

Fig. 5: Schematics of the mechanics of crack propagation. The
upper part of the cleaved multilayer partially shields the glue
layer from the singular field at the crack tip (see~\cite{Wei}).

\newpage
\section*{Table and Figures}
\begin{center}
\begin{tabular}{|c|c|c|}
\hline & multilayer & interfacial toughness (J/m$^2$)\\ \hline
  1 & Glass / \Sit // Ag / ZnO & \textcolor{red}{0.8, 0.8, 0.9}\\\hline
  2 & Glass / \Sit / ZnO // Ag / ZnO & 1.0, 1.1, 1.3 \\\hline
  3 & Glass / ZnO // Ag / ZnO & 1.4, 1.6\\ \hline
  4 & Glass / \TiOd / Ag // ZnO & 1.5, 1.9, 2,8 \\ \hline
  5 & Glass / \SnOd\ / Ag // ZnO & 2.4, 2.5\\ \hline
\end{tabular}
\end{center}

\newpage

\newpage

\newpage

\newpage

\newpage

\end{document}